# La Terra come osservatorio astronomico: la correzione al raggio solare medio nell'eclissi anulare del 22 settembre 2006

Costantino Sigismondi
ICRA e La Sapienza, Università di Roma    sigismondi@icra.it

**Sommario:** l'osservazione di un'eclissi di Sole dalla fascia di centralità, e ancor meglio, dal suo bordo, consente la misura del diametro solare con un'accuratezza fino ad una parte su 100.000.
Si presentano i dati preliminari dell'eclissi anulare del 22 settembre 2006 osservata da Kourou, Guyana francese e la loro calibrazione.
**Abstract:** the observation of a central eclipse from the umbral band and, even better, from its edge allows to measure the solar diameter with an accuracy up to a part over 100000. The preliminary data of the annular eclipse of 22 september 2006 observed from Kourou, French Guyane and their calibration are presented.

**Introduzione storica:**
Usare la dimensione e la forma della Terra per comparare osservazioni dello stesso corpo celeste e dedurne le sue dimensioni e distanze è un metodo che in Astronomia è stato inventato già alcuni secoli prima di Cristo, e di recente è stato "esportato" anche ad altre regioni dello spettro elettromagnetico come per le osservazioni VLBI di radiosorgenti.
Eratostene nel periodo ellenistico alessandrino, ed Halley nella seconda metà del seicento, sono i personaggi che restano legati a questi metodi per antonomasia.
Direttore della Biblioteca di Alessandria Eratostene (~272-192 a.C.) è famoso per il suo metodo per la misura della circonferenza della Terra di 250.000 stadi. Posidonio (135-51 a. C.) riporta anche il suo metodo per la misura del diametro angolare del Sole. A Syene (l'attuale Assuan) un palo verticale non getta ombra al mezzogiorno del solstizio d'Estate in una regione ampia 300 stadi, il diametro angolare del Sole spazza perciò una frazione pari a 300/250.000 di angolo giro.

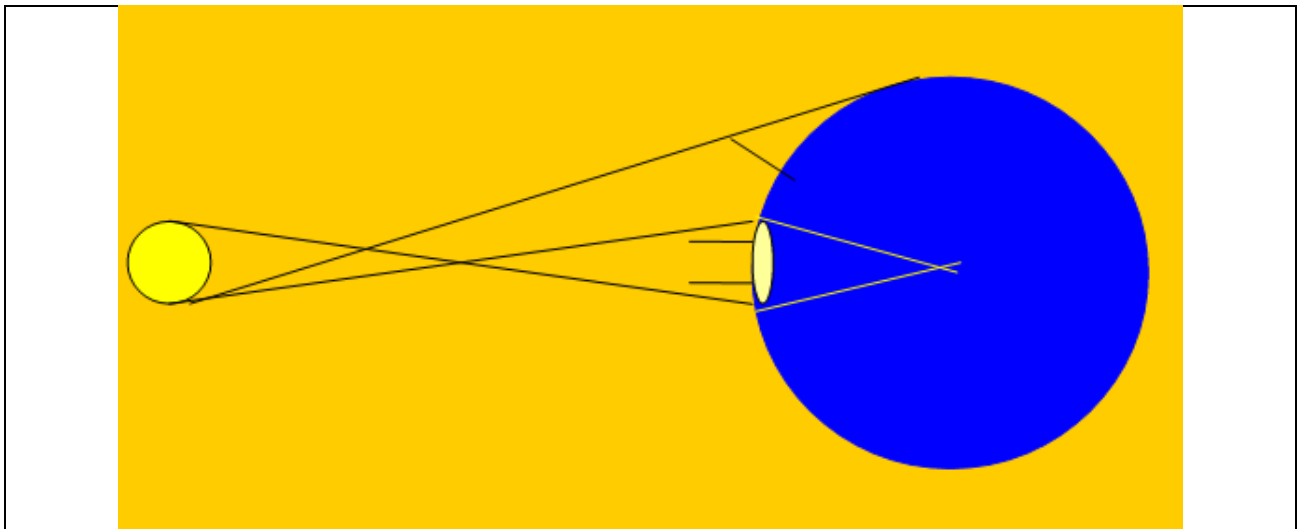

***Fig. 1*** *L'area in chiaro è la zona dove un palo non getta ombre durante il Solstizio d'Estate.*
*Sono stati rappresentati 2 pali che il Sole può avvolgere con la sua luce grazie al fatto di non essere una sorgente luminosa puntiforme. Un terzo palo decentrato da quell'area getta ombre via via più lunghe all'allontanarsi del Sole dal suo zenith. Distanze e dimensioni dei corpi qui rappresentati non sono in scala, ma di valore solo esemplificativo: in realtà il diametro del Sole è circa 110 volte più grande di quello della Terra e ad una distanza da noi di circa 11800 diametri terrestri.*

Halley nel 1677, dopo aver osservato il transito di Mercurio sul Sole del 7 novembre, propose di utilizzare quello di Venere (che non sarebbe accaduto fino al 1761) per valutare l'Unità Astronomica, la distanza media Terra-Sole. Halley pubblicò il metodo nelle Philosophical Transactions (vol. 34, pag. 454 (1716)) quasi quarant'anni dopo.

Il metodo consiste essenzialmente nel comparare due osservazioni dello stesso fenomeno da luoghi molto distanti tra loro sulla Terra, in questo modo le due corde descritte dal pianeta sul disco del Sole risultano separate tra loro di una quantità proporzionale all'angolo sotto cui la distanza tra i due osservatori si vedrebbe da Venere (fig. 2).

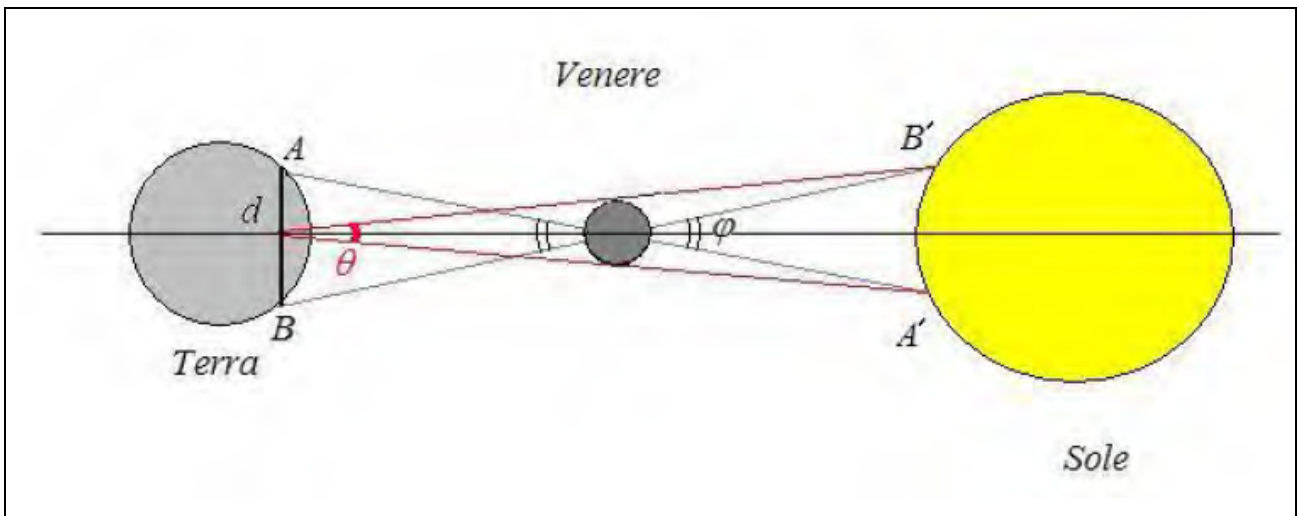

***Fig. 2*** *schema del Transito di Venere: la distanza tra due osservatori A e B lontani tra loro si riflette nella distanza B'A' tra le due corde descritte da Venere davanti al Sole.*

Lo stesso tipo di misura si può fare comparando le durate dei transiti, proporzionali alla lunghezza delle corde stesse.
L'unità astronomica è essenziale per valutare le dimensioni reali del Sole, date quelle angolari.
E' noto che Anassagora (434 a. C.) riteneva che il Sole fosse una massa infuocata poco più grande della Grecia; ma il metodo per valutarne la reale distanza lo dobbiamo ad Aristarco di Samo (III sec. a. C.) aveva trovato che quando la Luna si presenta come una perfetta mezzaluna, l'angolo fra le visuali del Sole e della Luna è inferiore ad un angolo retto per un trentesimo di quadrante (3°), e per questo il Sole è 19 volte più distante della Luna dalla Terra (cfr. Castelnuovo et al., 1986). Oggi sappiamo che questo angolo è 20 volte più piccolo, e la conseguente distanza 20 volte maggiore di quanto ritenuto da Aristarco e più di 1200 volte quella valutata da Anassagora.
Il contributo di Halley non si ferma alla misura dell'Unità Astronomica; egli guidò la campagna osservativa della Royal Society per l'eclissi di Sole del 3 maggio 1715 che fu totale sull'Inghilterra. Grazie a questi dati oggi abbiamo la misura più accurata del diametro angolare del Sole di tre secoli orsono, entro una parte su 10000. Questo dato è essenziale per la valutazione delle variazioni secolari del diametro della nostra stella.
L'idea di Halley fu quella di associare ad ogni dato di durata dell'eclissi quello della posizione dell'osservatore sulla Terra, in modo da valutare le dimensioni dell'ombra lunare sulla superficie terrestre ed in particolare della posizione dei suoi bordi settentrionale e meridionale.
I risultati pubblicati sulle Philosophical Transactions della Royal Society of London nel 1717 hanno consentito di identificare entro 50 m le posizioni degli osservatori che si trovavano ai bordi (a Darrington e a Cranbrok), che furono spettatori di un'eclissi di durata quasi istantanea o addirittura di un Sole che si ridusse ad un punto brillante come Marte senza mai sparire del tutto.

Su una fascia di quasi 400 km, questi 50 m hanno consentito una precisione di 1 parte su 10000 nella determinazione del diametro Solare angolare di allora.

Halley raggiunse una precisione 20 volte quella ottenuta da Posidonio con il metodo di Eratostene. Va notato che Posidonio spesso riportava misure altrui (Stahl, 1983), e che era interessato più al metodo che al risultato, e la stima sulla precisione del suo metodo è ottenuta considerando ± 1 stadio (circa 160 m) l'incertezza sulle dimensioni dell'area (~48 km) con i pali senza ombre al mezzodì del 21 giugno.

Dobbiamo al Prof. E. W. Brown della Yale University la ripresa del metodo di Halley per la campagna osservativa dell'eclissi del 25 gennaio 1925, totale su New York, la sua eredità è stata raccolta dal prof. S. Sofia che a Yale continua gli studi sull'evoluzione secolare del diametro solare da eclissi insieme ad altri colleghi americani (D. Dunham, W. Warren, A. Fiala) e a chi scrive.

Rispetto al 1715 oggi si osservano i grani di Baily, che sono determinati dall'alternarsi delle montagne e delle valli del bordo lunare, quando questo si frappone tra noi ed il lembo del Sole.

Ciò aumenta il numero di fenomeni temporalmente determinabili, migliorando di un fattore $1/\sqrt{N}$ la precisione sulla misura del diametro solare basata solo sugli istanti di inizio e fine della totalità.

**Localizzazione geografica**: la fascia dove l'eclissi è centrale, cioè o anulare o totale, è una ristretta zona del globo, larga attorno ai 2-300 km. Collocandosi ai bordi della fascia, dove la Luna si muove tangente al lembo del Sole, è possibile osservare i grani di Baily per un tempo anche superiore alla massima durata della totalità, e su un arco anche superiore ai 180 gradi. Nel caso dell'eclissi totale in Egitto del 29 marzo 2006 la corona solare è stata visibile ad occhio nudo dal bordo per 8 minuti.

In prima approssimazione l'accuratezza relativa sulla misura del diametro solare dipende anche dalla precisione con cui è nota la posizione dell'osservatore rispetto alle dimensioni della fascia.

Con un GPS *Garmin II Plus* ho raggiunto la precisione di ± 2 m da confrontare con i 317 km di ampiezza della fascia (Espenak, 2005), quindi di una parte su 160.000.

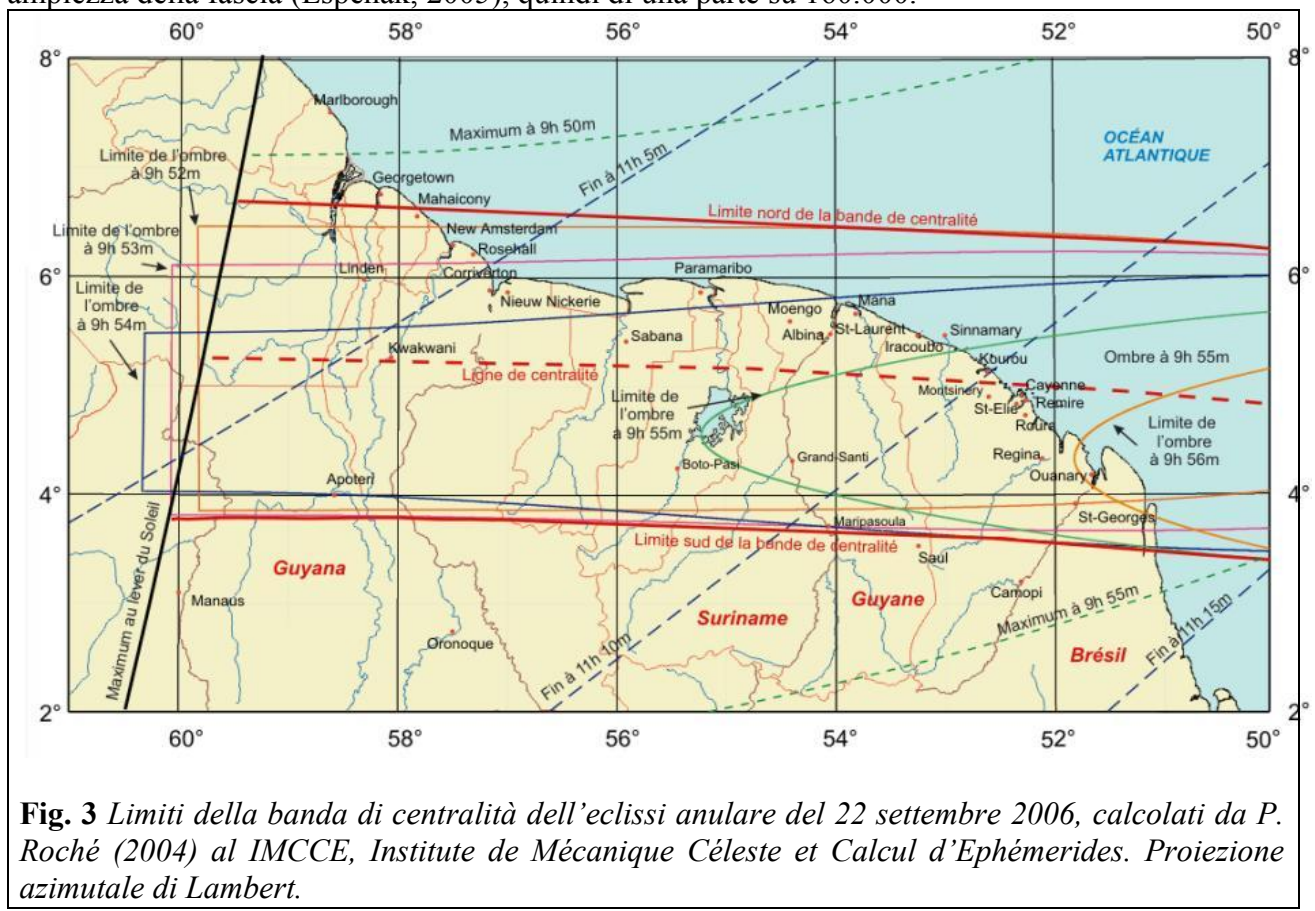

**Fig. 3** *Limiti della banda di centralità dell'eclissi anulare del 22 settembre 2006, calcolati da P. Roché (2004) al IMCCE, Institute de Mécanique Céleste et Calcul d'Ephémérides. Proiezione azimutale di Lambert.*

**Osservazione dei grani di Baily:** quando il bordo irregolare della Luna diventa tangente o quasi al lembo del Sole, le montagne lunari interrompono l'arco della fotosfera in diversi punti creando delle perle di luce. All'inizio della fase di anularità, quando il disco lunare entra nel disco solare che è più grande, queste perle si formano appena il Sole inizia a sorgere dalla valle lunare e spariscono fondendosi con le altre quando la fotosfera non è più interrotta dalle vette delle montagne lunari più alte. Al termine della fase di anularità sono le vette delle montagne più alte a toccare per prime la fotosfera e ad isolarne dei segmenti in cui il Sole tramonta alla base delle valli più profonde. In termini di velocità angolare la Luna avanza sul Sole a circa 0.5 arcsec/s e le montagne più alte arrivano ai 2 arcsec. Dunque sono circa 10 i secondi cruciali in cui tutti i grani di Baily si manifestano se osservati dal centro della fascia di anularità.

Gli eventi in studio sono quelli di apparizione (A) all'inizio dell'anularità e quelli di sparizione (S) alla fine. Degli istanti A ed S occorre conoscere l'istante del tempo universale coordinato UTC e l'angolo di posizione P.A. sul disco lunare.
Gli istanti degli eventi di fusione (merging) tra grani possono essere influenzati dal fenomeno della goccia nera (Pasachoof et al., 2004). Esso è dovuto alle distorsioni ottiche dello strumento combinate con il profilo radiale di luminosità del Sole rapidamente variabile, per cui con due strumenti diversi si registrano istanti diversi. Gli istanti di A ed S, invece, sono determinati da allineamenti geometrici al di fuori dell'atmosfera ed ogni possibile distorsione ottica in atmosfera o nel telescopio non modifica la natura del segnale che è di tipo luce SI / luce NO.
L'unica accortezza è quella di determinare la magnitudine limite dei grani di Baily osservabili, cioè il più piccolo numero di fotoni osservabile dal sistema telescopio-rivelatore, per il quale si può dire luce SI con un rapporto segnale – rumore pari almeno a 5.
Questo lavoro fino ad ora non era stato fatto in modo quantitativamente sufficiente (Sigismondi, 2006; Sigismondi e Oliva, 2006, Dunham 2005), e molte osservazioni del passato sono state ridotte senza tenere conto dell'effetto di diminuzione del diametro apparente dovuto ai filtri solari adoperati.

**Magnitudine limite dello strumento:**
Al fine di conoscere l'entità della riduzione del diametro solare dovuta all'uso di un filtro più o meno denso, ho scelto di osservare il Sole in eclissi con un sistema ottico di cui potessi misurare la trasmittanza, percentuale di energia trasmessa rispetto a quella incidente.
Poiché l'osservazione è stata fatta proiettando il Sole su uno schermo bianco su cui non cadessero i raggi diretti del Sole, quanto più l'immagine prodotta era grande tanto meno era il contrasto con il fondo. Dunque in questo particolare sistema ottico è il fondo cielo riflesso dallo schermo a determinare il livello di contrasto minimo per poter vedere i grani di Baily ad un determinato ingrandimento.
Test preliminari:
Una lampada di prova da 40 W, di diametro 3 cm era posta a 3 m 70 cm dall'obiettivo di un telescopio MEADE ETX 70, con lente obiettiva da 70 mm di diametro, con filtro YA2 che trasmette il 74.2% nel visuale. Usando l'oculare con prisma e lente raddrizzatori adoperati per l'eclissi, si otteneva sullo schermo un'immagine della lampada di 15 cm di diametro al limite di confusione per l'ambiente di osservazione utilizzato.
Un foro stenopeico di 2.1 mm di diametro produceva, nelle medesime condizioni di illuminazione, un'immagine di 7 mm di diametro su uno schermo a 65 cm.
Il telescopio con quell'oculare+lente raddrizzatrice ha una focale equivalente $F_{eq}$ che soddisfa all'equazione Dim=$F_{eq}\cdot\tan(\theta)$, dove Dim è il diametro dell'immagine e $\theta$ il suo diametro angolare (in radianti) visto dall'obiettivo del telescopio, cioè $\theta$=arctg(3/370)=0.465°=1672 arcsec=$8.1\cdot10^{-3}$ rad.
Risulta pertanto con il telescopio Dim=15 cm=$F_{eq}\cdot 8.1\cdot10^{-3}$ rad, da cui $F_{eq}$=18.5 m.

L'intensità luminosa di tale immagine va messa in relazione con quella prodotta dal foro stenopeico, la cui focale è 65 cm. Per il telescopio era a disposizione un'area di raccolta di 70 mm di diametro, mentre per il foro solo 2.1 mm, dunque $33.3^2 =1111$ volte tanto. L'intensità dell'immagine riflessa dallo schermo decresce inversamente proporzionale al quadrato della focale equivalente, e per il telescopio in rapporto al foro stenopeico abbiamo un valore di $(18.5/0.65)^2=810$ volte.

Poiché entrambe le immagini erano al limite della percettibilità significa che la trasmittanza del telescopio munito del filtro a tutta apertura YA2, laddove il semplice foro trasmette il 100% della radiazione incidente, è del 810/1111=72.9%. Il telescopio senza filtro trasmette il 98.25% della radiazione incidente.

**Trasmittanza dell'atmosfera:**

L'eclissi anulare del 22 settembre in Guyana Francese ha raggiunto la sua fase massima tra le 9:49 UT e le 9:55 UT, con l'altezza del Sole compresa tra 6°30' ed 8° sull'orizzonte Est.

A tali altezze l'estinzione dovuta all'atmosfera è piuttosto rilevante (rispettivamente 4.37 e 4.13 masse d'aria nel modello geometrico di atmosfera di Garstang, 1986).

La trasmittanza di una massa d'aria snella banda visuale V una massa d'aria trasmette l' 85.5% della radiazione (Barbieri, 1999).

Esperimenti condotti in Guyana, a Cayenne, all'alba e al tramonto ad Ostia sulla trasmittanza atmosferica nei pressi dell'orizzonte marino hanno evidenziato che l'estinzione atmosferica segue la legge $t=(0.855)^m$ fino a circa m=4.7 masse d'aria (circa 2° di altezza) per poi diventare dominata dalle brume dell'orizzonte.

Possiamo assumere che al momento dell'anularità, con l'altezza media del Sole corrispondente ad m=4.25 masse d'aria, la trasmittanza atmosferica fosse il 51.4%.

**Riduzione apparente del diametro solare: la funzione di oscuramento al bordo solare**

Per stimare quanto il Sole appaia più piccolo, quando si perdono alcune magnitudini nel sistema ottico di proiezione, occorre conoscere la curva di oscuramento al bordo solare (solar limb darkening function, SLDF).

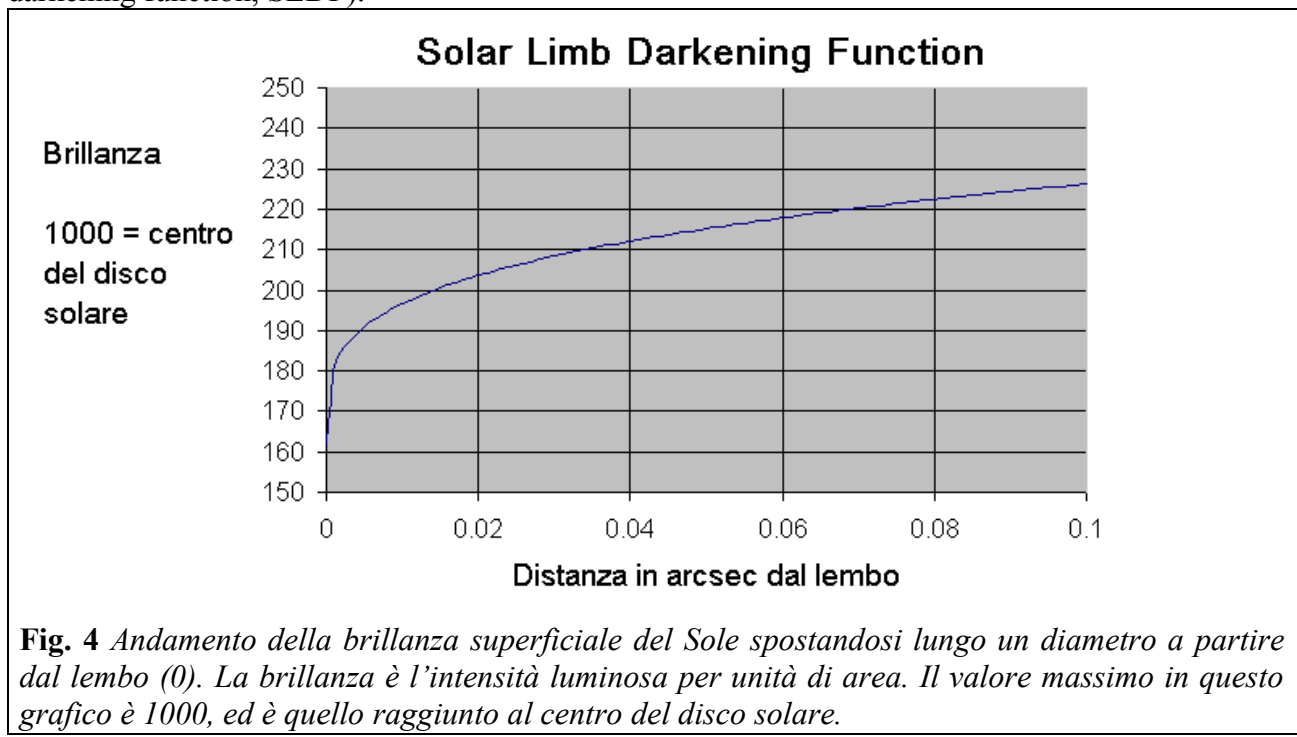

**Fig. 4** *Andamento della brillanza superficiale del Sole spostandosi lungo un diametro a partire dal lembo (0). La brillanza è l'intensità luminosa per unità di area. Il valore massimo in questo grafico è 1000, ed è quello raggiunto al centro del disco solare.*

Nota questa curva si calcola la luminosità di un bead in funzione della sua forma e profondità, cioè dell'area di fotosfera che la valle lunare ci lascia trasparire.
Consideriamo proprio le valli lunari che hanno determinato i beads osservati da Kourou, in Guyana Francese.

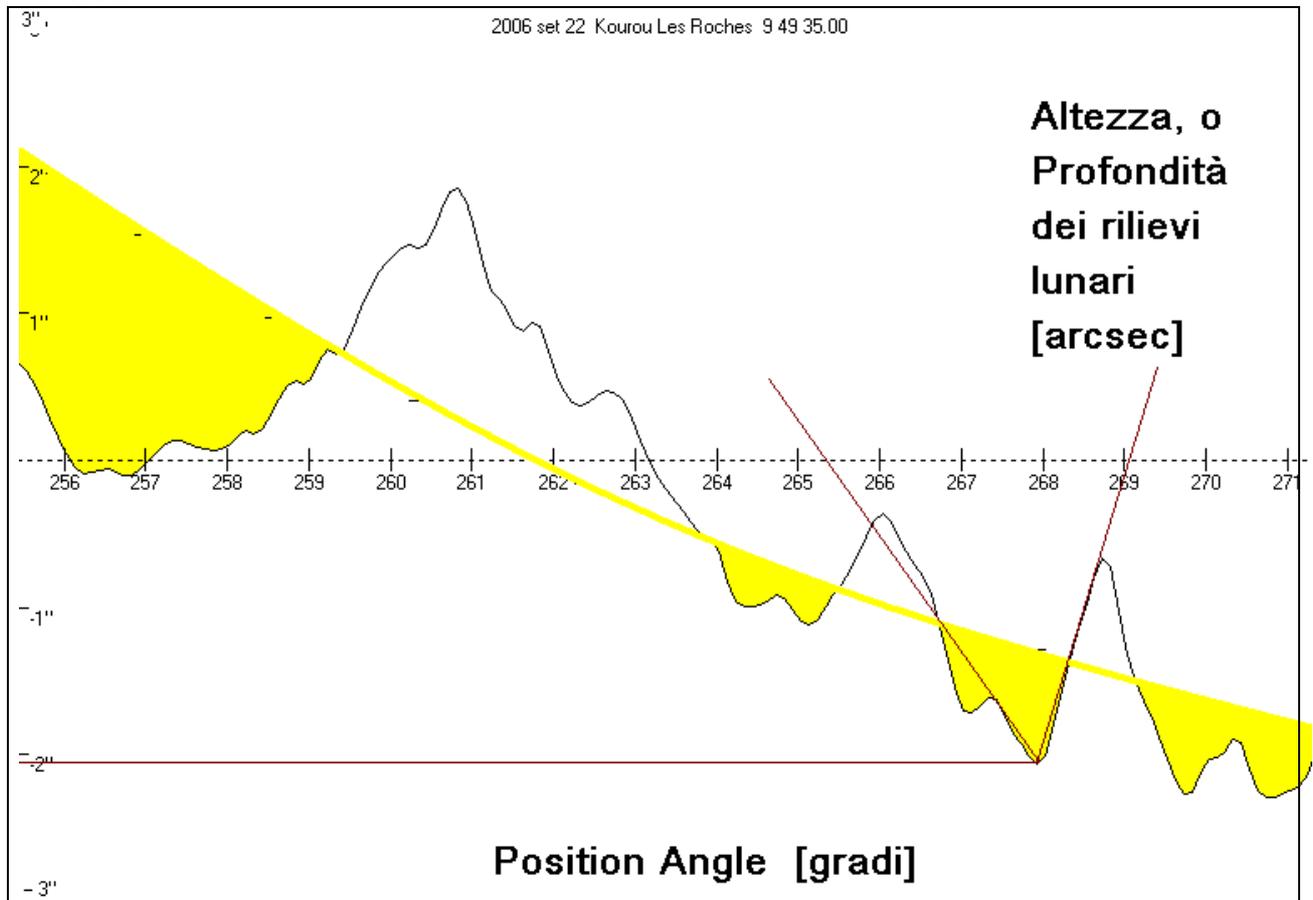

**Fig. 5** *I Baily Beads formati al secondo contatto, all'inizio dell'anularità. La profondità di h=2.1" corrisponde ad una base di 3.6° in Position Angle, che in arcosecondi corrisponde a b=55.4" (il diametro lunare valeva 1764" al momento dell'eclissi). La pendenza media vale h/(b/2)=0.076.*
*La simulazione è realizzata con il modulo Baily Bead del programma Winoccult 3.1 di David Herald.*

Con questo valore della pendenza l'area di un grano di Baily come funzione della sua profondità h vale $A=hb/2$, ma $h/0.076=(b/2)$, quindi $A=h^2/0.076$ arcsec².
Moltiplicando A per la brillanza media al lembo solare corrispondente alla profondità h si ottiene la luminosità del bead.

**Contrasto dell'immagine e luminosità limite del più piccolo grano di Baily osservabile:**
L'eclissi anulare del 22 settembre 2006 è stata osservata in proiezione su schermo.
Lo schermo essendo posto all'aria aperta prendeva luce parzialmente dal fondo cielo.
Quanto più il fondo cielo è illuminato quanto più luminosi devono essere i grani di Baily per risultare visibili. L'effetto del fondo cielo è quello di tagliare la parte esterna, meno brillante, del disco solare, offuscandola con la sua luce diffusa, dando così luogo ad una sottostima del diametro della fotosfera.

Stimiamo di seguito il più debole grano di Baily osservabile nelle condizioni di illuminazione della foto seguente.

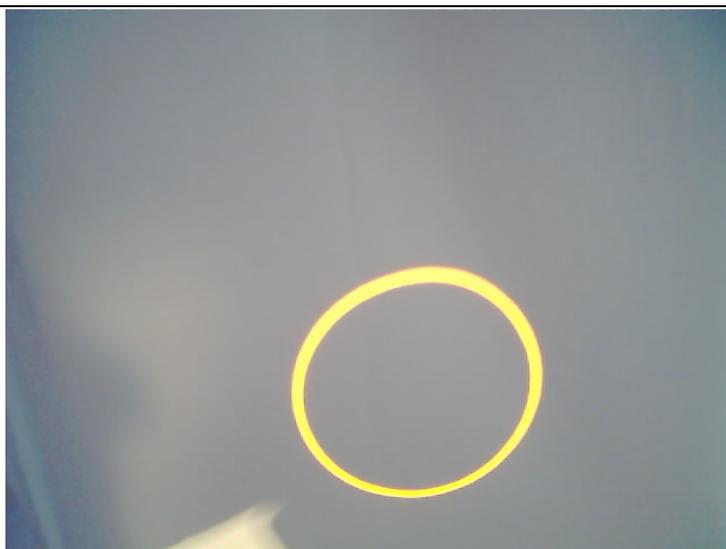

**Fig. 5** *Immagine dell'eclissi anulare in Guyana sullo stesso schermo dove sono state effettuate le riprese video dei Baily beads. Fase centrale dell'eclissi.*

Dall'analisi di questa immagine risulta un diametro di 204 pixel, corrispondenti a 1912 arcsec. Ogni pixel ha una dimensione di 9.4 arcsec.

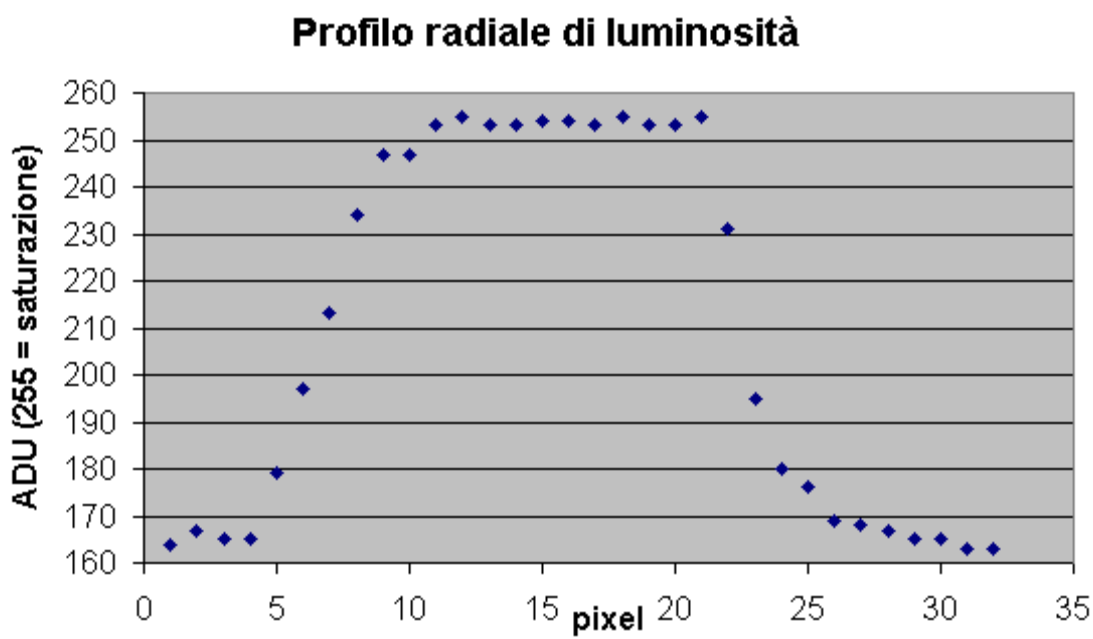

**Fig. 6** *Sezione radiale dell'anello (arco sinistro) della foto precedente, con l'intensità luminosa in ciascun pixel. Per l'estrazione dei dati di luminosità in unità adu dalla foto digitale si ringrazia Davide Troise. Analog-to-digital unit, ADU, è un numero che rappresenta l'ouput della CCD, il rivelatore di immagini della camera.*

Muovendosi lungo un diametro ed intercettando l'anello, la luminosità dei pixel cresce dal fondo (~ 163 ADU) alla saturazione (255 ADU) in 5 pixel, cioè 47 arcsec, come risulta dalla figura 6.

Secondo la SLDF a 47 arcsec dal lembo la luminosità per unità di area (brillanza) è 560/1000 del valore al centro del disco solare. L'area complessiva di un pixel vale 9.4 x 9.4 = 88.36 arcsec quadrati.

Ogni singolo pixel di 88.36 arcsec quadrati a brillanza 560/1000 del valore al centro del Sole, ridotta dei fattori 0.514 (trasmittanza atmosfera) moltiplicato per 0.729 (trasmittanza telescopio+filtro), corrisponde ad una quantità di fotoni registrata pari a 255 ADU.

Il rumore R è definito come la varianza del fondo cielo e risulta R=0.95 ADU. Una sorgente è identificabile come tale e non si confonde con le fluttuazioni del fondo se il suo segnale S>5R, cioè S>4.75 ADU.

Ora un bead profondo h sottende un'area di fotosfera $A=h^2/0.076$ alla brillanza media di SLDF(h), e la sua luminosità è ridotta del fattore 0.375 per la trasparenza di atmosfera e telescopio combinate. Tabulando i valori da 0 a 10 centesimi di arcosecondo otteniamo

| Profondità del grano di Baily [arcsec] | Area [arcsec²] | SLDF (h) [1000=centro] | Luminosità osservata con 37.5% trasmittanza [ADU] |
|---|---|---|---|
| Fotosfera | 88.36 | 560 | 255 |
| 0 | 0.000 | 162.0 | 0.00 |
| 0.1 | 0.011 | 226.2 | 0.15 |
| 0.2 | 0.047 | 239.6 | 0.65 |
| 0.3 | 0.110 | 248.9 | 1.52 |
| 0.4 | 0.202 | 256.2 | 2.78 |
| 0.5 | 0.323 | 262.3 | 4.45 |
| 0.6 | 0.475 | 267.6 | 6.53 |

**Tabella 1**. *Calcolo della luminosità (ADU) registrata da un grano di Baily durante l'eclissi anulare del 22 settembre 2006 in funzione della sua profondità. Il rivelatore di riferimento è la camera fotografica Philips Keyring 008, usata per la foto in fig. 5.*

Esaminando la Tabella 1 si vede che il grano di Baily raggiunge la luminosità limite di 4.75 ADU quando esso è profondo poco più di 0.51 arcsec. Un grano meno profondo non si vedrebbe, perché risulterebbe confuso con le fluttuazioni del fondo.

Poiché nell'analisi temporale dei Baily beads è stata usata una videocamera SONY da 800000 pixel dove un raggio del Sole occupava un intero lato dell'immagine (4:3) di circa 1000 pixel, abbiamo avuto una risoluzione spaziale almeno 10 volte superiore a quella della foto qui presentata, che ha consentito l'identificazione certa di 13 beads confrontando il video con le simulazioni del programma Winoccult 3.1 di David Herald.

La migliore risoluzione spaziale per la videocamera corrisponde anche ad una capacità di integrazione di luminosità 100 volte inferiori a parità di tempo di esposizione. Inoltre la macchina fotografica Philips Keyring 008 ha un tempo di esposizione di 1/100 s mentre la videocamera la ha di 1/25 s, con un guadagno in tempo di integrazione di ulteriori 4 volte.
La minima intensità rivelabile diventa 400 volte inferiore con la videocamera. Il contrasto 20 volte migliore. Questo nuovo livello di contrasto, se manteniamo per uniformità di discorso le stesse unità ADU della Keyring 008, corrisponderebbe a 0.238 ADU, e sarebbe raggiunto a 0.12 arcsec di profondità per il grano di Baily più debole osservabile.

Poiché la luminosità limite è stata stimata sull'immagine già proiettata dell'eclissi, questa tiene conto sia della trasmittanza del telescopio filtrato, che di quella dell'atmosfera, che dell'influenza del fondo cielo sullo schermo di proiezione.

Sappiamo che questi tre fattori possono cambiare per un'altra eclissi osservata con altri strumenti. In Egitto, il 29 marzo 2006, il Sole era ad oltre 60° sull'orizzonte al momento della totalità (Sole 1.6 volte più luminoso), con il cielo molto più scuro dell'eclissi anulare (fondo molto più basso, oltre 100 volte meno intenso). Considerando stavolta 0.0238 ADU (a causa del nuovo livello di fondo) come flusso limite rivelabile da quella videocamera per proiezione dello stesso telescopio, questo flusso veniva raggiunto a poco meno di 0.04 arcsec di profondità.

Quindi assumiamo che nell'eclissi anulare del 22 settembre 2006 il sistema telescopio+atmosfera e proiezione abbiano ridotto il raggio percepibile di 0.12 arcsec, mentre nell'eclissi totale del 29 marzo 2006 questo valore non superi gli 0.04 arcsec.

**Stima della correzione al raggio solare medio con Winoccult:**

La lista dei Baily beads identificati con il tempo di apparizione (a) / sparizione (s) ed il loro position angle sul lembo lunare (letto in senso antiorario, cioè verso Est, a partire da Nord) è riportata nella tabella 2.

| ORA UTC | evento | P.A. |
|---|---|---|
| 30.9 | a | 256.3 |
| 34.9 | a | 267.9 |
| 35 | a | 269.8 |
| 35.5 | a | 270.8 |
| 35.8 | a | 272.8 |
| 35.8 | a | 282.4 |
| 36 | a | 284.5 |
| | | |
| 19.7 | s | 97.4 |
| 20.5 | s | 105.5 |
| 20.5 | s | 107 |
| 21.6 | s | 116.2 |
| 25 | s | 85.8 |
| 26.4 | s | 122.3 |
| a: secondi dopo 9:49 UTC | | |
| d: secondi dopo 9:55 UTC | | |

**Tabella 2**. *Baily Beads identificati nell'eclissi anulare del 22 settembre 2006, osservata da Kourou, Guyana Francese, circa 150 metri a Nord della Tour Dreyfus, nel luogo chiamato Les Rochers, su una delle rocce paleozoiche presso la riva del mare. Le coordinate del punto di osservazione erano:*
*Latitudine 5° 9' 42.6" Nord*
*Longitudine 52° 37' 41.5" Ovest*
*Altitudine 3 m s.l.m.*

Il valore del raggio del Sole consistente con il timing assoluto di questi 13 Baily Beads così come si ottiene inserendo i dati nel programma Winoccult corrisponde a 0.04±0.04 arcsec rispetto al valore medio ad 1 Unità Astronomica di 959.63 arcsec. Aggiungendo 0.12 arcsec, per compensare la riduzione di raggio apparente di 0.12 arcsec nelle immagini dell'eclissi anulare, abbiamo che il raggio solare il 22 settembre 2006 risultava 0.16±0.04 arcsec più grande del suo valore medio.

Questo è anche il valore di riferimento del raggio solare al minimo di attività solare del ciclo 23, per quanto riguardo il metodo delle eclissi. Non ce ne saranno altre misurabili attorno a questo minimo.

L'incertezza statistica di ±0.04 arcsec corrisponde ad un'incertezza relativa di 1 parte su 50000 sul diametro solare. Questo è uno dei risultati più accurati ottenuti con le eclissi.

Poiché si ritiene che al minimo di attività il diametro raggiunga il suo massimo valore, questo massimo è di poco superiore al valor medio degli ultimi 50 anni ottenuto con vari metodi. Ciò è indizio del fatto che su base secolare il diametro del Sole si stia riducendo.

**Bibliografia:**

ROCHÉ, P., www.imcce.fr , *Eclipse annulaire de Soleil du 22 septembre 2006* (2004).

ESPENAK, F., http://sunearth.gsfc.nasa.gov/eclipse/OH/path/Path2006.html#2006Sep22A (2005).

PASACHOFF, J., SCHNEIDER, G. and L. GOLUB, *The black-drop effect explained*, Proc. IAU Colloquium 196, D.W. Kurtz & G.E. Bromage, eds., Cambridge University Press (2004).

SIGISMONDI, C*., Solar radius in 2005 and 2006 eclipses*, Solar Activity and its Magnetic Origin, Proceedings of the 233rd Symposium of the International Astronomical Union held in Cairo, Egypt, March 31 - April 4, 2006, Edited by Volker Bothmer; Ahmed Abdel Hady. Cambridge: Cambridge University Press, 2006., pp.519-520 (2006).

SIGISMONDI, C., OLIVA, P., *Astrometry and Relativity II: Solar Grazes & Solar Diameter,* Journal of the Korean Physical Society, vol. 49, pp.840-844 (2006).

DUNHAM, D. W. et al., *Accuracy of Solar Radius Determinations from Solar Eclipse Observations, and Comparison with SOHO Data*, 2005 SORCE Science Meeting September 14-16, Durango, Colorado (2005).

GARSTANG, R. H., *Model for Artificial Night-Sky Illumination*, Pub. A. S. P. 98, p. 364 (1986).

BARBIERI, C., *Lezioni di Astronomia*, Zanichelli, Bologna, 1999.

CASTELNUOVO, E., GORI GIORGI, C., VALENTI, D., *La Scienza, Trigonometria*, La Nuova Italia, Firenze, (1986).

STAHL, W., *La Scienza dei Romani*, Laterza, Bari, (1983).

THUILLIER, G., SOFIA, S. et al., Past, present and future measurements of the solar diameter" Advances in Space Research 35, p.329–340 (2005).

HERALD, D., Winoccult 3.1, http://www.lunar-occultations.com/iota/occult3.htm (2006).